\documentclass[12pt]{article}
\usepackage{amsmath,amssymb}

\begin{document}

\title{A cohomological construction of integrable hierarchies of 
hydrodynamic type}
\author{Paolo Lorenzoni\\
Dipartimento di Matematica e Applicazioni\\
Universit\`a di Milano-Bicocca\\
Via R. Cozzi 53, I-20126 Milano, Italy\\
paolo.lorenzoni@unimib.it\\[2ex]
Franco Magri\\
Dipartimento di Matematica e Applicazioni\\
Universit\`a di Milano-Bicocca\\
Via R. Cozzi 53, I-20126 Milano, Italy\\
franco.magri@unimib.it
}
\maketitle
\vspace{.2 cm}
\begin{abstract}
\noindent We explain how to use the theory of bidifferential ideals 
to construct integrable hierarchies of hydrodynamic type
\end{abstract}

\section{Introduction}
In this note we show how to use the theory of bidifferential 
ideals to solve a classical problem of the theory of first-order partial 
differential equations of hydrodynamic type
\begin{eqnarray*}
\frac{\partial u}{\partial t}=A(u)\frac{\partial u}{\partial x}
\end{eqnarray*}
in two independent variables $(x,t)$ and $n$ dependent variables 
$u=(u^1,...,u^n)$ with periodic boundary conditions. 
 The problem is to construct an infinite family 
\begin{eqnarray*}
\frac{\partial u}{\partial t_i}=A_i(u)\frac{\partial u}{\partial x}
\end{eqnarray*}
of equations of this type whose flows commute in pair, 
so that each equation can be considered as defining a symmetry of all 
the remaining equations. Simultaneously the problem requires to 
construct an infinite family of functionals 
\begin{eqnarray*}
I_l[u]=\int_{S^1}k_l(u)dx
\end{eqnarray*}
which are constants of motion of all the previous flows, so that
\begin{eqnarray*}
\frac{\partial I_l}{\partial t_k}=0
\end{eqnarray*}
for all values of the indexes $l,k\in \mathbb{N}$.

As usual  we consider a system of  hydrodynamic type as a vector field
 on a loop space $\hat{\cal 
F}$, whose points 
are 
the $C^{\infty}$-maps from the circle to a base manifold $\cal F$ 
endowed 
with coordinates $(u^1,...,u^n)$. On this manifold the problem requires
 to construct 
an infinite family of pairwise commuting vector fields. A simple way for 
dealing with this problem is to consider a linear operator 
$N:T\hat{\cal F}\rightarrow T\hat{\cal F}$, called a recursion 
operator, and 
to define the family of commuting vector fields $X_j$ through the 
recursive relation
\begin{eqnarray*}
X_{j+1}(u)=N_u X_j(u).
\end{eqnarray*}
It is known that, if the torsion of $N$ vanishes, and if the first 
vector field $X_0$ of the hierarchy is chosen in such a way that the Lie 
derivative of $N$ along $X_0$ vanishes, then all the vector fields 
$X_j$ commute in pair.

Although conceptualy simple, the previous scheme has several practical 
drawbacks, the main of which is that the recursion operator $N$ is 
often a nonlocal operator (that is an integro-differential operator), so 
that it is extremely difficult to give the theory a sound formal basis. 
Moreover, there is no reason to restrict our attention to such simple 
recursion relations as that written before. That would entail a 
noticeable loss of interesting examples.

In this note we show how to amend the theory of recursion operator of 
both drawbacks (at least in the setting of PDEs of hydrodynamic type). 
Our theory rests on two simple ideas. The first is that the recursion 
operator $N:T\hat{\cal F}\rightarrow T\hat{\cal F}$, on the loop 
space, may 
be conveniently replaced by a much simpler recursion operator 
$L:T{\cal F} \to T{\cal F}$ on the base manifold $\cal F$. The second 
idea is that 
the simple recursion relation of Lenard's type is just a particular 
instance of a general class of recursion relations provided by the 
theory of ``prolongation of bidifferential ideals''. Despite the exotic 
language, the new scheme of iteration is very simple, and it  
may be of use  for more general class of equations than those of 
hydrodynamic type.

\vspace{.5 cm}
{\bf Aknowledgments}. The authors thank A. Shabat for a 
discussion on the weakly nonlinear systems which was the starting point 
of the present work, and B. Dubrovin and M. Pedroni for useful comments. 

\section{Bidifferential ideals and hydrodynamic type hierarchies}

A tensor field $L:T{\cal F}\rightarrow T{\cal F}$, of type $(1,1)$ on a manifold $\cal F$, 
of dimension $n$, is torsionless if the following identity
\begin{eqnarray*}
[LX,LY]-L[LX,Y]-L[X,LY]+L^2[X,Y]=0
\end{eqnarray*}
is verified for any pair of vector fields $X$ and $Y$ on $\cal F$, the 
bracket denoting the standard commutator of vector fields. One of the 
reasons to consider a torsionless tensor field $L$ is that it defines, 
according to the theory of graded derivations of Fr\"{o}licher-Nijenhuis 
(see \cite{FN}), a differential operator $d_L$, of degree 1 and type 
$d$, on the Grasmann algebra of differential forms on $\cal F$, verifying the 
fundamental conditions
\begin{eqnarray*}
d\cdot d_L+d_L\cdot d=0\hspace{2cm}d_L^2=0.
\end{eqnarray*}
On functions and 1-forms this derivation is defined by the following 
equations 
\begin{eqnarray*}
&&d_L f(X)=df(LX)\\
&&d_L\alpha(X,Y)=L_{LX}(\alpha(Y))-L_{LY}(\alpha(X))-\alpha([X,Y]_L),
\end{eqnarray*}
where
\begin{eqnarray*} 
[X,Y]_L=[LX,Y]+[X,LY]-L[X,Y].
\end{eqnarray*}
Since the pair of differential 
operators $d$ and $d_L$ define a double cohomological complex, 
 one may introduce the concept of bidifferential ideal of 
forms. 

\newtheorem{de}{Definition}
\begin{de}
A bidifferential ideal $\mathfrak{I}$ is an ideal of differential forms 
on 
$\cal F$ which is closed with respect to the action of both $d$ and $d_L$:
\begin{eqnarray*}
d(\mathfrak{I})\subset\mathfrak{I}\hspace{3 
cm}d_L(\mathfrak{I})\subset\mathfrak{I}
\end{eqnarray*}
\end{de}

Let us make this concept concrete by a simple example. Assume that 
$\mathfrak{I}$ has rank 1, and therefore that is generated by a single 
1-form $\alpha$. The above equations then require that there exist two 
1-forms $\lambda$ and $\mu$ (called the Frobenius multipliers) such that
\begin{eqnarray*}
d\alpha=\lambda\wedge\alpha\hspace{3
cm}d_L\alpha=\mu\wedge\alpha.
\end{eqnarray*}
In virtue of the first condition and on account of the Frobenius 
theorem, one may assume, without loss of generality, that the 1-form 
$\alpha$ is exact: $\alpha=dh$. Therefore any bidifferential ideal of 
rank 1 is defined by a function $h$ for which there exists a 1-form 
$\mu$ such that
\begin{eqnarray*}
dd_L h=\mu\wedge dh.
\end{eqnarray*}
Due to the basic identities 
$d^2=0$ and $d_L^2=0$, the 1-form $\mu$ must verify the pair of conditions
\begin{eqnarray*}
d\mu\wedge dh=0\hspace{3 cm}d_L\mu\wedge dh=0.
\end{eqnarray*}
A special class of ideals is selected by imposing that the multiplier 
$\mu$ obeys the stronger conditions
\begin{eqnarray*}
d\mu=0\hspace{3 cm}d_L\mu=0.
\end{eqnarray*}
Even more stringently, one may assume that $\mu$ is an exact one form, 
$\mu=da$, and that its potential $a$ verifies the remaining condition
\begin{equation}
\label{ddla}
dd_L a=0.
\end{equation}
By this process of successive restrictions, one thus selects a special 
class of bidifferential ideals that are  called flat.

\begin{de}
A flat bidifferential ideal $\mathfrak{I}$, of rank 1, on a manifold $\cal F$ 
endowed with a torsionless tensor field $L:T\cal F\rightarrow T\cal F$, is the ideal 
of forms generated by the differential $dh$ of a function $h:\cal F\rightarrow \cal F$ 
 obeying the condition
\begin{eqnarray}
\label{ddlh}
dd_L h=da\wedge dh
\end{eqnarray}
with respect to a conformal factor $a$ which in turn verifies
 the cohomological condition (\ref{ddla}).
\end{de}
Since any solution of (\ref{ddla}) is also a solution of the associated equation (\ref{ddlh}) 
there exists a special class of bidifferential ideals generated by a single function $a$.  

The main purpose of the present note is to prove that an integrable hierarchy 
of hydrodynamic type on $\hat{\cal F}$ is associated with any flat bidifferential ideal,
 of rank 1, on $\cal F$. Probably this correspondence may be extended to higher
 rank flat bidifferential ideals on $\cal F$ (to be defined in a similar way),
 but in this note we shall stick, for 
simplicity, to the rank 1 case. The key to understand
 the relation between bidifferential ideals on $\cal F$ 
 and integrable hierarchies of hydrodynamic type on $\hat{\cal F}$ is the process of 
prolongation  of bidifferential ideals which is presently introduced.

Let us consider again an arbitrary bidifferential ideals $\mathfrak{I}$ on $\cal F$ 
which is generated by the 1-forms $(\alpha_1,...,\alpha_p)$, and let us denote by 
$\mathfrak{I'}$ the ideal which is generated by the same forms and by their iterations 
$(L^*\alpha_1,...,L^*\alpha_p)$. One can prove that $\mathfrak{I'}$
 is still a bidifferential ideal. So the family of all bidifferential ideals on $\cal F$ is closed 
under the action of $L$. Since $\mathfrak{I'}$ clearly includes the original ideal 
$\mathfrak{I}$, it is natural to refer to this process
 as the process of prolongation of bidifferential ideals. In 
particular, if $\mathfrak{I}$ is flat, one can show that also $\mathfrak{I'}$ is flat. 
This means that $\mathfrak{I'}$ is generated by a family of exact 1-forms $dh_j$, $j=1,...,q$ 
obeying the conditions 
\begin{eqnarray*}
dd_L h_i=\sum_{k=1}^q da^k_j\wedge dh_k
\end{eqnarray*}
with conformal factors $a^k_j$ verifying the ``flatness conditions''
\begin{eqnarray*}
dd_L a^k_j=\sum_{l=1}^q da^l_j\wedge da^k_l.
\end{eqnarray*}
The proof of these claims in full generality is not difficult
 but is irrelevant 
for the purpose of the present note. So we limit ourselves to detail the process 
of prolongation of bidiferrential ideals in the particular case of rank 1 ideals (\cite{M2}).
\newtheorem{pro}{Proposition}
\begin{pro}
Let $\mathfrak{I}$ be a rank 1 flat bidifferential ideal on $\cal F$,
 generated by a function $h$ satisfying the condition 
(\ref{ddlh}) with a conformal factor $a$ satisfying
 the flatness condition (\ref{ddla}). Then the successive prolongations 
$\mathfrak{I'}$, $\mathfrak{I''}$,... of $\mathfrak{I}$
 are flat bidifferential ideals of rank 2,3,... which are generated by the possibly 
 infinite sequence of functions $(h_0,h_1,h_2,...)$ recursively defined by the relations
\begin{eqnarray*}
&&dh_{k+1}=d_L h_k-a_k dh_0\\
&&da_{k+1}=d_L a_k-a_k da_0
\end{eqnarray*}
where $a_0=a$ and $h_0=h$.
\end{pro}
{\bf Proof}

Start from equations (\ref{ddlh}) and (\ref{ddla}) written 
in the form
\begin{eqnarray*}
&&d(d_L h_0-a_0 dh_0)=0\\
&&d(d_L a_0-a_0 da_0)=0.
\end{eqnarray*}
Locally there exists a pair of functions $(h_1,a_1)$ such that
\begin{eqnarray*}
&&dh_1=d_L h_0-a_0 dh_0\\
&&da_1=d_L a_0-a_0 da_0.
\end{eqnarray*}
Let us apply the differential $d_L$ to both sides of these equations. 
One obtains
\begin{eqnarray*}
dd_L h_1 &=& d_L a_0\wedge dh_0-a_0dd_L h_0\\
         &=& d_L a_0\wedge dh_0-a_0da_0\wedge dh_0\\
         &=& da_1\wedge dh_0
\end{eqnarray*}
and
\begin{eqnarray*}
dd_L a_1=d_L a_0\wedge da_0=da_1\wedge da_0.
\end{eqnarray*}
Let us now proceeds by induction, assuming that there exists a couple of functions 
 $(h_k,a_k)$ such that 
\begin{eqnarray*}
&&dd_L h_k-da_k\wedge dh_0=0\\
&&dd_L a_k-da_k\wedge da_0=0.
\end{eqnarray*}
These conditions imply the existence of a new pair of functions  $(h_{k+1},a_{k+1})$  
such that
\begin{eqnarray*}
&&dh_{k+1}=d_L h_k-a_k dh_0\\
&&da_{k+1}=d_L a_k-a_k da_0.
\end{eqnarray*}
By applying $d_L$ to both sides, one readily obtains
\begin{eqnarray*}
&&dd_L h_{k+1}=da_{k+1}\wedge dh_0\\
&&dd_L a_{k+1}=da_{k+1}\wedge da_0,
\end{eqnarray*}
so the iterative process starts again. The form of the iteration
 immediately shows that the differentials $dh_k$ generate the successive prolongations
 of the original flat bidifferential ideal $\mathfrak{I}$ generated 
by the pair of functions $(h_0,a_0)$.
 The proof that the prolonged ideals are still flat is omitted, since it is of no 
interest for the present note.
\begin{flushright}
$\Box$
\end{flushright}

The previous result clearly points out the relation between bidifferential
 ideals and chains. The functions defining  a flat bidifferential ideal
 on $\cal F$ may be used as starting elements of   
appropriate recurrence relations defining, at each step, a new set of functions.
 These recurrence relations translate on functions the process
 of prolongation of bidifferential ideals. The results 
of each step of the iterative procedure define the  members of the chain 
associated with the original  bidifferential ideal.
 The missing part of the present theory is to show that the chain of bidifferential 
ideal on $\cal F$, just described, may be converted into a chain of 
commuting vector fields of hydrodynamic type on the loop space $\hat{\cal F}$ related to $\cal F$.

 We have all the tools required to perform this last step except two. 
The first is the class of tensor fields $M_k$, of type $(1,1)$,
 recursively defined by 
\begin{eqnarray*}
M_{k+1}=M_k L-a_k E,
\end{eqnarray*}
starting from the identity $E$. The first elements of this family are
\begin{eqnarray*}
&&M_0=E\\
&&M_1=L-a_0 E\\
&&M_{2}=L^2-a_0 L-a_1 E
\end{eqnarray*}
and so on. They naturally arise in the present picture, since they are 
the linear operators relating the differentials ($da_k,dh_k$) of the iterated functions 
to the initial differentials ($da_0,dh_0$) through the relations
\begin{eqnarray*}
&&da_k=M^*_k da_0\\
&&dh_k=M^*_k dh_0.
\end{eqnarray*}
The second missing element is a new chain of functions $(k_0,k_1,k_2,...)$ which may be built 
from the functions $h_k$ and $a_k$ coming from the process of prolongation of 
bidifferential ideals. One may  define the new functions either by specifying 
how they are related to $(h_k,a_k)$ by means of relations of the form
\begin{eqnarray*}
&&k_0=h_0\\
&&k_1=h_1+a_0 h_0\\
&&k_2=h_2+a_0 h_1+(a_1+a_0^2)h_0\\
&&k_3=h_3+a_0 h_2+(a_1+a_0^2)h_1+(a_2+2a_1 a_0+a_0^3)\\
&&................
\end{eqnarray*}
or, more conveniently, by the recurrence relation 
\begin{equation}
\label{dk+}
d k_{l+1}=d_L k_{l}+k_l da_0
\end{equation}
starting from $k_0=h_0$. One may easily check that this recurrence relation 
provides, at each step, an exact 1-form, thus allowing to compute a new 
iterated function. Moreover, by applying $d_L$ to both sides of (\ref{dk+}) we obtain the condition
\begin{eqnarray*}
dd_L k_{l+1}=d_L k_l\wedge da_0=dk_{l+1}\wedge da_0
\end{eqnarray*}
which means that the differentials $dk_0$, $dk_1$, $dk_2$,... define a chain of flat 
bidifferential ideals having the same conformal factor $a_0$. 

We can now state the main 
result of the note.
\begin{pro}
The chain of vector fields of hydrodynamic type defined on $\hat{\cal F}$ by 
 the equations
\begin{eqnarray*}
\frac{\partial u}{\partial t_j}=M_j u_x
\end{eqnarray*}
commute in pair, and the chain of functionals
\begin{eqnarray*}
I_l=\int_{s^1}k_l(u)dx
\end{eqnarray*}
are the related integrals. So any flat bidifferential ideal of 
rank 1 on $\cal F$ allows to construct an integrable hierarchy of 
hydrodynamic type on the loop space $\hat{\cal F}$ associated with $\cal F$.
\end{pro}
{\bf Proof}

In order to prove the first claim we recall (see \cite{PSS}) 
that two vector fields $u_t=A u_x$ and $u_{\tau}=B u_x$ commute if and only if 
the tensor fields of type $(1,1)$ $A$ and $B$ commute as linear operator
\begin{eqnarray*}
&&A\cdot B-B\cdot A=0,
\end{eqnarray*}
and verify the differential condition
\begin{eqnarray}
\label{cc}
&&[AX,BX]-A[X,BX]-B[AX,X]=0
\end{eqnarray}
for any vector field $X$ on ${\cal F}$. In our case the first condition is
 obviously satisfied. It remains to prove the condition (\ref{cc}). To this end, 
it is suitable to introduce the symbol 
\begin{eqnarray}
&&\{AX,BX\}=[AX,BX]-A[X,BX]-B[AX,X]
\end{eqnarray}
to condense in the formula
\begin{eqnarray}
\label{iden}
\{M_{k+1}X,M_{l+1}X\}=L\cdot\{M_{k}X,M_{l+1}X\}+L\cdot\{M_{k+1}X,M_{l}X\}
-L^2\cdot\{M_{k}X,M_{l}X\}
\end{eqnarray}
a useful identity relating the four tensor field $(M_k,M_{k+1},M_l,M_{l+1})$. 
 This identity is proved in Appendix , by exploiting the vanishing of 
the torsion of $L$ and the recurrence relations on the conformal factors $a_k$. 
The above identity holds for all integers $k,l\in \mathbb{N}$. Notice that for $M_0=E$ 
 one has
\begin{eqnarray*}
\{M_0 X,M_l X\}=0,
\end{eqnarray*}
for all $l\in \mathbb{N}$, meaning that the first vector field $\hat{X}_0=M_0 u_x$ commute 
with all the remaining vector fields $\hat{X}_l=M_l u_x$ of the hierarchy:  
\begin{eqnarray*}
[\hat{X}_0,\hat{X}_l]=0.
\end{eqnarray*} 
Assume to have proved that, for a certain $k$, the vector field $\hat{X}_k=M_k u_x$
 commute with all the vector fields of the hierarchy:
\begin{eqnarray*}
[\hat{X}_k,\hat{X}_l]=0\hspace{1 cm}\forall l\in\mathbb{N}.
\end{eqnarray*}
This means to assume that
\begin{eqnarray*}
\{M_k X,M_l X\}=0\hspace{1 cm}\forall l\in\mathbb{N}.
\end{eqnarray*}
Then the identity (\ref{iden}) shows that
\begin{eqnarray*}
\{M_{k+1} X,M_l X\}=L\{M_{k+1} X,M_{l-1} X\}
\end{eqnarray*}
and, by iteration, that
\begin{eqnarray*}
\{M_{k+1} X,M_l X\}=L^l\{M_{k+1} X,M_{0} X\}=0
\end{eqnarray*}
Therefore, if $\hat{X}_k$ commute with the hierarchy, also $\hat{X}_{k+1}$ 
commute with the hierarchy. This remark ends the proof of the first part of 
the theorem.

\vspace{.5 cm}
In order to prove the second part, it is sufficient to 
notice the following recursive relations on the derivatives of the functionals $I_j$:
\begin{eqnarray*}
\hat{X}_{k+1}(I_j) &=& \int_{S^1}<dk_j,L\hat{X}_k -a_k \hat{X}_0>dx=\\
             &=& \int_{S^1}(<d_{L}k_j,\hat{X}_k>-<dk_j,a_k \hat{X}_0>dx=\\
             &=& \int_{S^1}(<dk_{j+1}-k_j da_0,\hat{X}_k>-<dk_j,a_k \hat{X}_0>)dx=\\
             &=& \int_{S^1}<dk_{j+1},\hat{X}_k>dx-\int_{S^1}(<k_jda_k+a_k dk_j,\hat{X}_0>=\\
             &=& \hat{X}_k(I_{j+1})-\int_{S^1}<d(k_j a_k),\hat{X}_0>dx=\\
             &=& \hat{X}_k(I_{j+1})-\int_{S^1}\frac{d}{dx}(k_j a_k)dx=\\
             &=& \hat{X}_k(I_{j+1})
\end{eqnarray*}
Therefore
\begin{eqnarray*}
\hat{X}_K(I_j)=\hat{X}_{k+1}(I_{j+1})=......=\hat{X}_0(I_{j+k})=0
\end{eqnarray*}
since $\hat{X}_0(I_{l})=0$ for any $l$.
\begin{flushright}
$\Box$
\end{flushright}
\section{The diagonal case}

Let us restrict our attention to the case where $L:T\cal F\rightarrow T\cal F$
 has real and distinct eigenvalues. On account of the vanishing of the torsion of $L$, 
this assumption entails the existence of a system of local coordinates $(q^1,...,q^n)$ 
 such that
\begin{eqnarray*}
L^*dq^i=f^i(q^i)dq^i
\end{eqnarray*}
The basic condition (\ref{ddla}), starting the iterative scheme of bidifferential ideals, 
 takes the simpler form
\begin{eqnarray*}
\partial_i\partial_j a_0=0,
\end{eqnarray*}
showing that the most general admissible conformal factor $a_0$ is a sum 
of functions of a single coordinate: 
\begin{eqnarray*}
a_0=\sum_i g^i(q^i).
\end{eqnarray*}
It follows that the first equation of the hierarchy of equations of hydrodynamic type 
generated by $a_0$ has the form
\begin{eqnarray}
\label{ds}
q^i_t=(f^i(q)-a_0)q^i_x.
\end{eqnarray}
These equations have been previously considered by Pavlov, who obtained them through the study,
 in his language, of the finite-component reductions of the infinite momentum chain
\begin{eqnarray*}
\partial_t c_k=\partial_x c_{k+1}-c_1\partial_x c_k\hspace{1
cm}k=0,\pm 1,\pm 2,...\hspace{1cm}.
\end{eqnarray*}
One can thus say that the construction of the previous section explains,
 in the case of a diagonalizable recursion operator $L$, the geometric meaning of the 
Pavlov's reductions.

It is known that the system (\ref{ds}) is semi-Hamiltonian, that is 
 the  characteristic speeds $v^i(q)=f^i-a_0$ verify
 the ``complete integrability'' conditions 
\begin{equation}
\label{sh}
\partial_j\left(\frac{\partial_k v^i}{v^k-v^i}\right)=
\partial_k\left(\frac{\partial_j v^i}{v^j-v^i}\right)\hspace{1 
cm}\forall i\ne j\ne k\ne i.
\end{equation}
It is also known that the functional   
 $I[u]=\int h(u)dx$ is a first integral of a semi-Hamiltonian system $q^i_t=v^i(q)q^i_x$ if and only if
 the density $h$ satisfies the following system (see \cite{ts}) 
\begin{eqnarray}
\label{cs}
\partial_i\partial_j h-\frac{\partial_j v^i}{v^j-v^i}\partial_i h-
\frac{\partial_i v^j}{v^i-v^j}\partial_j h=0\hspace{1 cm}i\ne j.
\end{eqnarray}
In order to link the present approach to the existing literature, we observe that 
 in the present case, where $v^i=f^i-a_0$, the equations (\ref{sh}) and (\ref{cs}) read:
\begin{eqnarray*}
&&\partial_i\partial_j a_0=0\\
&&(f^i-f^j)\frac{\partial^2 h}{\partial q^i\partial q^j}=\frac{\partial a_0}{\partial q^i}
\frac{\partial h}{\partial q^j}-\frac{\partial a_0}{\partial q^j}
\frac{\partial h}{\partial q^i}.
\end{eqnarray*}
They are nothing else than the coordinate form of the basic
 equations (\ref{ddla}) and (\ref{ddlh}):
\begin{eqnarray*}
&&dd_L a_0=0\\
&&dd_L h=da_0\wedge dh
\end{eqnarray*}
This remark provides a new insight on the meaning of these equations. The thorough  
study of the non diagonal case is outsides the limits of the present note.
\section{Appendix}
This section is devoted to the proof of the identitity 
\begin{eqnarray*}
\{M_{k+1}X,M_{l+1}X\}=L\cdot\{M_{k}X,M_{l+1}X\}+L\cdot\{M_{k+1}X,M_{l}X\}
-L^2\cdot\{M_{k}X,M_{l}X\}
\end{eqnarray*}
By using the relation $M_{k+1}=M_k L-a_k E$, we obtain
\begin{eqnarray*}
&&\{M_{k+1}X,M_{l+1}X\}=\\
&&\left([LM_k X,LM_l X]-[a_kX,LM_l X]-[LM_k X,a_lX]+[a_k X, a_l X]\right)+\\
&&+\left(-LM_k [X,M_{l+1}X]+a_k[X,M_{l+1}X]\right)+\left(-LM_l [M_{k+1}X,X]+a_l[M_{k+1}X,X]\right)
\end{eqnarray*}
Let us compute the sum of the first terms in each bracket:
\begin{eqnarray*}
&&[LM_k X,LM_l X]-LM_k [X,M_{l+1}X]-LM_l [M_{k+1}X,X]=\\
&&[LM_k X,LM_l X]+L\cdot\{M_k X,M_{l+1}X\}-L[M_k X,M_{l+1}X]+LM_{l+1}[M_k X,X]+\\
&&+L\cdot\{M_{k+1} X,M_{l}X\}-L[M_{k+1} X,M_{l}X]+LM_{k+1}[X,M_l X]=\\
&&L\cdot\left(\{M_k X,M_{l+1}X\}+\{M_{k+1} X,M_{l}X\}\right)
-L^2\cdot\{M_{k}X,M_{l}X\}-a_l L[M_k X,X]+\\
&&-a_k L[X,M_l X]+L^2[M_k X,M_l X]+[LM_k X,LM_l X]-L[M_k X,LM_l X]+\\
&&+L[M_k X,a_l X]-L[LM_k X,M_l X]+L[a_k X,M_l X]
\end{eqnarray*}  
Taking into account that (for the vanishing of the torsion of $L$)
\begin{eqnarray*}
[LM_k X,LM_l X]+L^2[M_k X,M_l X]-L[M_k X,LM_l X]-L[LM_k X,M_l X]=0,
\end{eqnarray*}
and that
\begin{eqnarray*}
&&-a_l L[M_k X,X]+L[M_k X,a_l X]=<M_k X,da_l>LX=<M_l M_k X,da_0>LX\\
&&-a_k L[X,M_l X]+L[a_k X,M_l X]=<M_l X,da_k>LX=<M_k M_l X,da_0>LX
\end{eqnarray*}
we obtain finally
\begin{eqnarray*}
&&[LM_k X,LM_l X]-LM_k [X,M_{l+1}X]-LM_l [M_{k+1}X,X]=\\
&&L\cdot\{M_{k}X,M_{l+1}X\}+L\cdot\{M_{k+1}X,M_{l}X\}
-L^2\cdot\{M_{k}X,M_{l}X\}
\end{eqnarray*}
It remains to prove that the sum of the remaining terms vanishes. But this follows 
from the following sequence of identities:
\begin{eqnarray*}
&&-[a_k X, LM_l X]-[LM_k X,a_l X]+[a_k X, a_l X]+a_k[X, M_{l+1}X]+a_l[M_kX,X]=\\
&&-[a_k X, M_{l+1} X]+a_k[X,M_{l+1} X]-[M_{k+1}X,a_l X]+a_l[M_{k+1} X,X]-[a_k X, a_l X]=\\
&&<M_{l+1}X,da_k>X-<M_{k+1}X,da_l>X+a_l<X,da_k>X-a_k<X,da_l>X=0 
\end{eqnarray*}

\end{document}